\documentclass[aps,pra,twocolumn,groupedaddress]{revtex4}

\usepackage{graphicx}
\usepackage{dcolumn}
\usepackage{bm}
\usepackage{amsmath}
\usepackage{cases}
\usepackage{color}
\usepackage{tabularx}
\newcolumntype{Y}{>{\centering\arraybackslash}X} 

\newcommand{\SI}[1]{\textcolor{black}{#1}}

\begin{document}


\title{Faraday waves in Bose--Einstein condensate: From instability to destabilization dynamics}



\author{Kasumi Okazaki$^1$}
\author{Junsik Han$^1$}
\author{Makoto Tsubota$^2$}
\affiliation{$^1$Department of Physics, Osaka City University, 3-3-138 Sugimoto, 558-8585 Osaka, Japan} 
\affiliation{$^2$Department of Physics \& Nambu Yoichiro Institute of Theoretical and Experimental Physics (NITEP) \& The OCU Advanced Research Institute for Natural Science and Technology (OCARINA), Osaka City University, 3-3-138 Sugimoto, Sumiyoshi-ku, Osaka 558-8585, Japan}

\date{\today}

\begin{abstract}
We numerically study the dynamics of Faraday waves in a pancake-shaped Bose--Einstein condensate (BEC) subject to periodic modulation of the interaction.
\SI{We solve the two-dimensional Gross--Pitaevskii equation with or without dissipation, and the obtained dynamics is much different between two cases.}
After the modulation starts \SI{without the dissipation}, Faraday waves appear.
By maintaining the modulation, the kinetic energy \SI{coming from} the density gradient causes quasiperiodic motion and increases monotonically.
\SI{Thereafter, the BEC enters the ''destabilization regime'', in which lots of collective modes are excited.}
In this regime, the dips in the density similar to dark solitons move around in the BEC, intersecting with each other and maintaining their shapes.
\SI{The time-averaged spectrum of the momentum distribution in this regime obeys a power law similar to wave turbulence.}
\SI{In the dissipative case, the simulation illustrates the characteristic dynamics of suppression and revival of the destabilization as well as Faraday waves.}
%
%
\SI{This is a nonequilibrium state sustained by the balance between the injection and the dissipation.}
\end{abstract}

\pacs{xxxx}

\maketitle


\section{Introduction}
Fluid mechanics has been studied in various fields of research such as turbulence \cite{David2015}, vortices \cite{Acheson1990, Batchelor1967}, hydrodynamic instability \cite{Chandrasekhar1981}, and pattern formations \cite{Cross1993}.
In pattern formation, as the control parameters increase, complex bifurcation and various nonequilibrium phenomena appear through nonlinear interaction.
Spatial pattern formation in a classical fluid is primarily categorized into three types: Taylor--Couette flow, Rayleigh--Benard convection, and parametric surface waves.
The representative example of the parametric surface waves is Faraday waves \cite{Faraday1831}, which are regular standing waves that appear on the surface of a vibrating fluid.
These waves are excited with twice of a natural frequency.
In 1831, Faraday waves were observed in an experiment conducted by \SI{oscillating vertically a container} containing fluids.
These dynamics are governed by an equation of pattern formations known as the Mathieu equation \cite{Benjamin1954}:
\begin{equation} \label{eq: Mathieu}
\frac{d^2x}{dt^2}+\Omega^2\left[1+\epsilon\cos\left(\omega t\right)\right]x=0,
\end{equation}
where $x$ represents the fluctuation from a stationary state, $\epsilon$ is the drive amplitude, $\Omega$ is the natural frequency, and $\omega$ is the drive frequency.
When the higher order of the fluctuations of surface waves can be ignored, the Euler equation reduces to a Mathieu equation \cite{Benjamin1954}.
The Floquet analysis of a Mathieu equation yields an infinite series of resonances at $\omega=2\Omega/n$ for integer $n$, and the first ($n=1$) resonance excites Faraday waves.
Several studies on Faraday waves have demonstrated solitons \cite{Wu1984}, pattern formations \cite{Edwards1993, Edwards1994, Ezerskii1985}, and space and time modulation \cite{Ezerskii1985, Ezerskii1990, Ezerskii1989}.
%

The quantum hydrodynamics of a Bose-–Einstein condensate (BEC) \cite{Tubota2013} have also been studied for topics such as quantum turbulence \cite{Madeira2020}, quantized vortices, quantum hydrodynamic instability, and pattern formations.
The dynamics of a BEC with a wave function $\psi$ are governed by the Gross–-Pitaevskii (GP) equation \cite{Pethick2008} as follows:
\begin{equation} \label{eq: GP}
(i-\gamma)\hbar \frac{\partial \psi}{\partial t} = -\frac{\hbar^2}{2m}\nabla^2\psi + V\psi + g|\psi|^2\psi,
\end{equation}
where $m$ is the mass of an atom, $V$ is the potential, and $g$ represents the interaction of atoms. 
We introduce the term $\gamma$ to describe the dissipation \cite{Choi1998, Tsubota2002}.
Faraday waves in a BEC are excited experimentally by modulation of the trapping potential \cite{Engels2007} or the interaction \cite{Nguyen2019}.
When $V=0$, substituting $\psi=\exp\left[-i\frac{\alpha}{\omega}\sin\left(2\omega t\right)\right]\left[1+w(t)\cos (kx)\right]$ in the GP equation (Eq. (\ref{eq: GP})) by modulating the interaction gives the Mathieu equation \cite{Staliunas2002}.
If a BEC is trapped by a cigar-shaped potential, forceful resonant waves similar to Faraday waves are experimentally excited with $\omega=2\omega_r$ along the longitudinal direction by the modulation of the potential along the tight direction or the modulation of the interaction \cite{Engels2007, Nguyen2019}.
These parametrically driven excited waves have been studied theoretically \cite{Nicolin2007, Balaz2014, Modugno2006}.
Although these situations in a BEC are different from the original Faraday waves in a classical fluid, such excited waves in a BEC can be considered as an analog to Faraday waves.
%

The Faraday waves of a BEC have been studied theoretically and experimentally.
Previous studies \cite{Engels2007, Nguyen2019} have reported that Faraday waves appear by modulating the trapping potential or the interaction.
In the first case of modulating the trapping potential \cite{Engels2007}, the experiments were conducted by periodic modulation of the transverse trapping frequencies of a cigar-shaped BEC.
Engels $et$ $al.$ observed the appearance of longitudinal Faraday waves with $\omega=2\omega_r$, where the trapping frequency $\omega_r$ corresponds to $\Omega$ (Eq. (\ref{eq: Mathieu})).
They also determined the instability of Faraday waves by strongly driving the transverse breathing mode.
In the second case, when the interaction is modulated in a cigar-shaped BEC \cite{Nguyen2019}, Nguyen $et$ $al.$ observed Faraday waves after the modulation time of $t_m=5$ ms, followed by a holding time of $t_h=20$ ms.
There have also been certain theoretical and numerical studies of Faraday waves \cite{Staliunas2002, Staliunas2004, Balaz2012, Kagan2001, Balaz2014, Nicolin2007, Modugno2006, Nath2010, Lakomy2012, Verma2017, Abdullaev2013}.
When $V=0$ \cite{Staliunas2002}, Staliunas $et$ $al.$ solved the GP equation \SI{for a BEC with the modulated} interaction.
The results indicated the evolutions of patterns and resonance tongues of parametric instability for conservative and dissipative systems.
They \cite{Staliunas2004} also indicated the appearance of a self-parametric instability in the case of one-dimensional cigar-shaped and two-dimensional disk-shaped BECs.
Furthermore, the study represents the dependency of the numerically determined threshold on the modulation frequency.
Balaz $et$ $al.$ \cite{Balaz2014} numerically calculated the time evolutions of the radially integrated longitudinal density profile by modulating the trapping potential in a collisionally inhomogeneous BEC.
The time evolutions indicated that the appearance of Faraday waves is accompanied by the excitation of the collective modes, and \SI{these waves force the condensate into a destabilized state.}
Faraday waves have been studied in two-component BECs \cite{Balaz2012, Maity2020} by Balaz $et$ $al.$ and Maity $et$ $al.$.
When the trapping potential is modulated \cite{Balaz2012}, the time evolutions from two initial states of dark-bright symbiotic pair states and segregated states were simulated by coupled GP equations.
As the result, Faraday waves were excited simultaneously in the two components.
%

In this study, we examine Faraday waves, instability, and destabilization regime by modulating the interaction in a two-dimensional anisotropic pancake-shaped BEC.
The cigar-shaped and isotropic pancake-shaped BEC have been studied previously \citep{Nguyen2019, Staliunas2002}, thus we consider the anisotropic pancake-shaped BEC, which should reveal the essence of the dynamics of the \SI{quasi one-dimensional} Faraday waves.
We assume that the \SI{ground-state} hyperfine level $|F=1, m_F=1\rangle$ of the $^7$Li atom and the s-wave scattering length of the atoms is controlled by Feshbach resonance.
We carry out three types of calculations.
One is by maintaining the modulation, the other is by turning the modulation off at $t_m=5$ ms without the dissipation $\gamma$, and the third one is maintaining the modulation along with the dissipation $\gamma$.
We numerically illustrate the appearance of Faraday waves in all the calculations for $\omega=2\omega_x$.
\SI{Furthermore, we reveal for the first time that the revival and suppression of these waves requires the dissipation.}
%
For the nondissipative case, after the first appearance of Faraday waves, the system enters the destabilization regime, \SI{which consists of the excitations of the collective modes and represents the periodic motion of the density dips.}
%

This paper is organized as follows.
In Sec. \ref{sec: Cal}, we describe our numerical scheme of a two-dimensional GP equation and the conditions of our calculations.
In Sec. \ref{sec: Ex of Fw}, we present our numerical results \SI{that indicate the excitation of Faraday waves and the instability.}
Thereafter, the BEC without the dissipation enters the destabilization regime.
The destabilization regime discussed in Sec. \ref{sec: Eme of non} consists of the excitations of collective modes and Faraday waves in the BEC with the dissipation repeats the expansion and suppression \SI{when the modulation of the interaction is maintained.}
Finally, in Sec. \ref{sec: Con}, we summarize the study.

\section{Calculation}  \label{sec: Cal}
We numerically solve the two-dimensional GP equation (Eq. (\ref{eq: GP})) by modulating the interaction parameter using the spectral method.
We consider a pancake-shaped BEC of $8\times 10^5$ $^7$Li atoms in a harmonic trap, $V=\frac{1}{2}m\left(x^2\omega_{x}^2+y^2\omega_{y}^2\right)$.
The trapping frequencies are $\omega_{x}=2\pi\times 512$ Hz and $\omega_{y}=2\pi\times 8$ Hz.
We consider a two-dimensional anisotropic BEC that is tightly confined and approximated by a Gaussian distribution in the $z$ direction.
The spatial mesh sizes in the $x$ and $y$ directions are $dx=0.025$ and $dy=0.025$, respectively.
\SI{The size of computational windows are about 20 $\mu$m along the $x$ direction and about 500 $\mu$m along the $y$ direction.}
\SI{We assume the periodic boundary condition.}
The interaction parameter is modulated using Feshbach resonance \cite{Inoue1998}.
The s-wave scattering length is controlled by the magnetic field \cite{Pethick2008}; thus, the interaction parameter can be expressed as follows:
\begin{equation} \label{eq: interaction energy}
g=\frac{\hbar^2}{m}a_{bg}\left(1+\frac{\Delta}{B(t)-B_\infty}\right), 
\end{equation}
where $a_{bg}=-24.5a_0$, $a_0$ is the Bohr radius, $\Delta=192.3$ G, and $B_\infty=736.8$ G.
The oscillation of the magnetic field is expressed as follows:
\begin{equation} \label{eq: modulation}
B(t)=\overline{B}+\Delta B\sin(\omega t),
\end{equation}
where $\overline{B}=577.4$ G is the mean magnetic field, $\Delta B$ is the amplitude, and $\omega$ is the modulation frequency.
To investigate the amplitudes $\Delta B$ and the modulation frequencies $\omega$ necessary for the appearance of Faraday waves, we calculate the GP equation using the various values of $\Delta B$ and $\omega$.
These frequencies correspond to the modes $n$ of the Mathieu equation (Eq. (\ref{eq: Mathieu})).
It is convenient to introduce dimensionless variables here:
\begin{equation} \label{eq: dimensionless}
t=\frac{1}{\omega_y}\tilde{t}, x=a_y\tilde{x}, y=a_y\tilde{y}, \psi=\sqrt{N}\frac{\tilde{\psi}}{a_y},
\end{equation}
where $a_y=\sqrt{\hbar/m\omega_y}$ is the length along the $y$-axis and $N$ is the total particle number expressed as follows:
\begin{equation} \label{eq: dimensionless}
N=\int_{}{}\int_{}{} |\psi (x,y)|^2 dxdy.
\end{equation}
The GP equation (Eq. (\ref{eq: GP})) is reduced to a dimensionless form: 
\begin{equation} \label{eq: dimensionless GP}
(i-\gamma) \frac{\partial \tilde{\psi}}{\partial \tilde{t}} = -\frac{1}{2}\tilde{\nabla^2}\tilde{\psi} + \tilde{V}\tilde{\psi} + \tilde{g}|\tilde{\psi}|^2\tilde{\psi}.
\end{equation}
The dimensionless potential and interaction take the following forms: $\tilde{V}=1/2\left(\tilde{x}^2\omega_x^2/\omega_y^2+\tilde{y}^2\right)$ and
\begin{equation} \label{eq: dimensionless g}
\tilde{g}=2\sqrt{2\pi}N\frac{a_{bg}}{a_x}\left(1+\frac{\Delta}{B(t)-B_\infty}\right),
\end{equation}
where $a_x=\sqrt{\hbar/m\omega_x}$ is the length along the $x$-axis.
The dimensionless symbol tilde is omitted below for simplicity.
%

To evaluate how Faraday waves appear depending on the modulation time $t_m$ and the dissipation, we calculate the GP equation for three cases.
For $\gamma=0$, the first case is to maintain the modulation and the second is turning the modulation off at $t_{m}=5$ ms.
The third case is to maintain the modulation along with the dissipation $\gamma$.
We calculate the third case because a system with dissipation is realistic and demonstrates a qualitatively different dynamics than one without dissipation.
%

Next, we calculate the kinetic energy.
The total energy $E_{tot}$ of a BEC is expressed as follows:
\begin{equation} \label{eq: Etot}
E_{tot}=\int_{}^{} \left[\frac{1}{2}|\nabla\psi|^2+V|\psi|^2+g|\psi|^4\right] dxdy,
\end{equation}
where the first, second, and third terms on the right-hand side are the kinetic, potential, and interaction energies, respectively.
For a wave function expressed as $\psi=|\psi|e^{i\theta}$, the kinetic energy \SI{density} is calculated as follows:
\begin{equation} \label{eq: Ek}
|\nabla\psi|^2=\left|\nabla|\psi|\right|^2+|\psi|^2\left(\nabla\theta\right)^2,
\end{equation}
where the first and second terms represent the contributions of the density gradient and the superfluid velocity $v_s=\hbar/m\left(\nabla\theta\right)$ to the kinetic energy.

\section{Excitation and dynamics of Faraday waves} \label{sec: Ex of Fw}
In this section, we present the results of the three cases mentioned above.
We also discuss the instability against the modulation and excitation of Faraday waves.
We discuss two main results in this section.
First, the calculations for $\omega=2\omega_x$ and $1$ G $<\Delta B<9$ G indicate the appearance of Faraday waves.
\SI{Interestingly, Faraday waves appear even if $n$ of the resonance condition $\omega=2\Omega /n$ of a Mathieu equation is a fractional value.}
The second result is that the dynamics changes to the destabilization regime, such as excitations of collective modes similar to wave turbulence after the excitation of Faraday modes in the dissipationless case.
However, the calculation along with the dissipation indicates that Faraday waves appear and disappear with a cycle of hundreds of milliseconds without entering the destabilization regime.
The analysis of the destabilization regime is described in Sec. \ref{sec: Eme of non}, and this section describes only the results of Faraday waves and instability.

\subsection{Appearance of Faraday waves without dissipation ($\gamma=0$)}
\begin{figure}[t] 
   \begin{center} 
       \includegraphics [width=1\columnwidth] {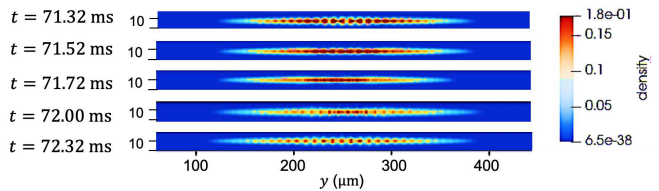}
       \caption{Dynamics of Faraday waves during $t=71.32--72.32$ ms in BEC. These figures represent density images of Faraday waves in harmonic trap. Driving frequency $\omega=1024\times 2\pi$ Hz and amplitude $\Delta B=5$ G. This calculation maintains modulation without $\gamma$.}
       \label{fig: g1_7_fw_dy}
    \end{center}
\end{figure}

When we continue to apply the modulation without dissipation,  Faraday waves that have regularly spaced patterns are excited after $t\simeq 68$ ms.
Figure \ref{fig: g1_7_fw_dy} displays the typical dynamics of the Faraday wave density.
These waves are excited at $t=71.32$ ms and start to suppression at $t=71.52$ ms.
After the suppression, the excited waves reappear at $t =72.32$ ms.
This suppression and revival for a few milliseconds are repeated.
\begin{figure}[t] 
   \begin{center} 
       \includegraphics [width=1\columnwidth] {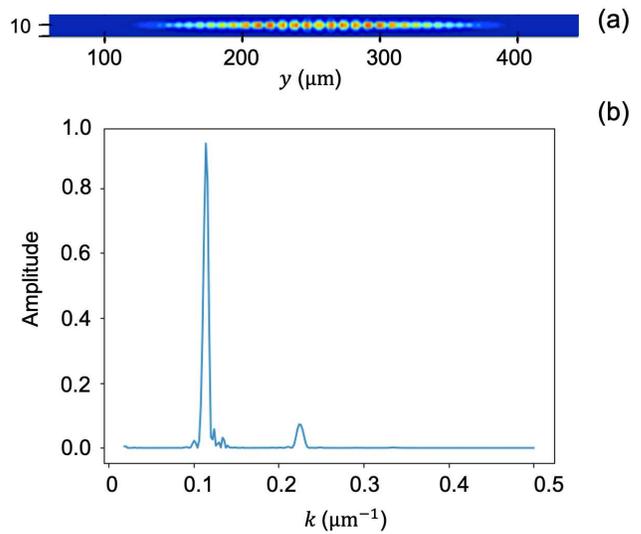}
       \caption{Appearance of Faraday waves at $t\simeq 80$ ms. Driving frequency $\omega=1024\times 2\pi$ Hz and amplitude $\Delta B=5$ G. This calculation maintains modulation without $\gamma$. (a) Density image of Faraday waves and (b) FT of one-dimensional density.}
       \label{fig: g1_7_fw}
    \end{center}
\end{figure}
Faraday waves are characterized by the single dominant peak of a Fourier transform (FT).
The spectrum is calculated by integrating $|\psi|^2$ along the $x$-axis:
\begin{equation} \label{eq: integrate of FFT}
|\psi_{1D}(y,t)|^2=\int_{-\infty}^{\infty} |\psi(x,y,t)|^2 dx.
\end{equation}
We apply the FT to the one-dimensional density $|\psi_{1D}|^2$ and obtain the spectrum of regularly spaced patterns.
A typical image of Faraday patterns is displayed in Fig. \ref{fig: g1_7_fw}.
Figures \ref{fig: g1_7_fw} (a) and (b) display the density and FT spectrum, respectively, when these waves appear.
Figure \ref{fig: g1_7_fw} (b) indicates that the spectrum is characterized by a single peak $k\simeq 0.12$ $\rm{\mu}$m$^{-1}$ corresponding to a spatial period of the density $\lambda\simeq 8.70$ $\rm{\mu}$m.
\SI{This wave number is estimated theoretically from the stability of the Mathieu equation.}
\SI{The GP equation with modulating the interaction is reduced to the Mathieu equation \cite{Staliunas2002, Nicolin2011}.}
\SI{When we apply the well-known scenario of the instability of the Mathieu equation to our case \cite{Mc1951}, the wave number of the excited Faraday waves is found to be $k=0.14$ $\mu$m$^{-1}$, which agrees well with our numerical result $k\simeq 0.12$ $\rm{\mu}$m$^{-1}$.}
This result is similar to the experimental result \cite{Nguyen2019}, thus for the elongated BEC, the Faraday waves appear as shown in Fig. \ref{fig: g1_7_fw}.
The beginning of the animation \cite{Sup1, Sup2} represents the time development of Faraday waves.
%

%
\begin{figure}[b] 
   \begin{center} 
       \includegraphics [width=1\columnwidth] {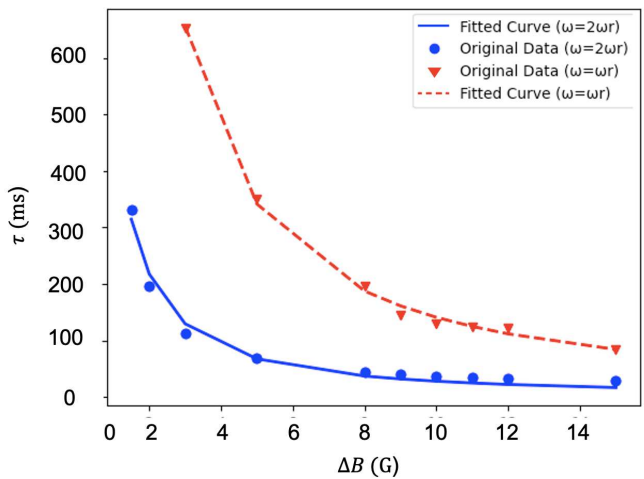}
       \caption{Instability onset times $\tau$ plotted as a function of $\Delta B$. Blue circles and red triangles represent results of calculations with modulation for $\omega=2\omega_{x}$ and $\omega=\omega_x$ without $\gamma$, respectively. Blue solid and red dashed lines are fitted curves.}
       \label{fig: ins_t_B}
    \end{center}
\end{figure}
We also observe that the instability onset time $\tau$ depends on $\omega$ and $\Delta B$.
The instability onset time $\tau$ refers to the time at which the excited density waves such as Faraday modes and other modes appear for the first time.
The onset time for Faraday waves decreases with $\Delta B$, as depicted in Fig. \ref{fig: ins_t_B}, and has different values for each $\omega$.
That is, the instability appears quickly as the injected energy depending on $\Delta B$ increases.
When $\omega=2\omega_x/n$ ($1 < n$), the density waves of mode $n$ appear.
This onset time is generally delayed compared to that of the Faraday waves and this resonance is the strongest.
%

%
\begin{figure}[t] 
   \begin{center} 
       \includegraphics [width=1\columnwidth] {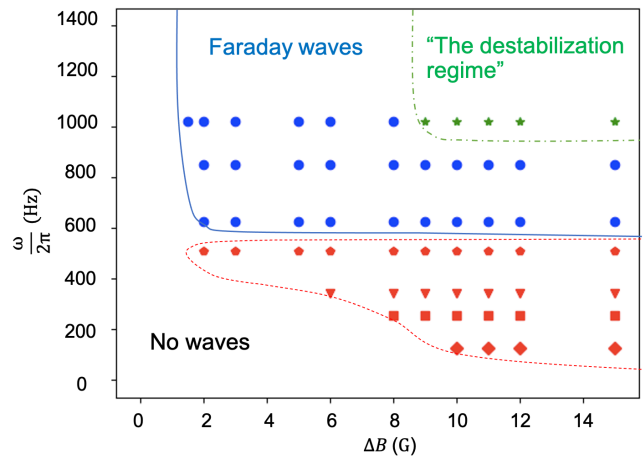}
       \caption{Phase diagram of instability for $\omega$ and $\Delta B$. We calculate GP equation for $\omega=2\omega_{x}$ ($n=1$), $\omega=\frac{5}{3}\omega_x$ ($n=\frac{6}{5}$), $\omega=\frac{3}{2}\omega_x$ ($n=\frac{4}{3}$), $\omega=\omega_{x}$ ($n=2$), $\omega=\frac{2}{3}\omega_{x}$ ($n=3$),  $\omega=\frac{1}{2}\omega_{x}$ ($n=4$), and $\omega=\frac{1}{4}\omega_{x}$ ($n=8$) by maintaining modulation without $\gamma$. Trapping frequency $\omega_x=2\pi\times 512$ Hz. Blue circles and red pentagons represent appearance of Faraday modes ($n=1$) and resonance modes ($n=2$), respectively. Red triangles, red squares, red diamonds, and green stars represent modes $n=3$, $n=4$, and $n=8$ and destabilization regime, respectively. Regions surrounded by solid blue line, dashed green line, and dotted red line represent excitation of Faraday waves, destabilization regime, and density waves of modes of $n\neq 1$, respectively. Other region depicts no waves.}
       \label{fig: ob}
    \end{center}
\end{figure}
Figure \ref{fig: ob} illustrates \SI{what kind of density waves appears at first depending on $\omega$ and $\Delta B$.}
Faraday waves are not clearly observed even at $\omega=2\omega_{x}$ for the modulation amplitude of $\Delta B \leq 1$ G or $\Delta B\geq 9$ G.
For $\Delta B\geq 9$ G, Faraday modes appear along with other collective modes; however, it quickly casts the system into the destabilization regime.
%
%
When $\Delta B \leq 1$ G, the system is almost stationary because the modulation amplitude is too small.
Hence, Faraday waves clearly appear for $1$ G $<\Delta B< 9$ G.
\SI{Please note that the destabilization regime appears even for $1$ G $<\Delta B< 9$ G after the appearance of Faraday waves.}
If the modulation frequency $\omega$ and amplitude $\Delta B$ are too small, these waves are not observed.
Thus, in subsequent calculations, we primarily assume the conditions of $\omega=2\omega_x$ and $\Delta B=5$ G.
%

When we turn off the modulation at $t_m=5$ ms, Faraday waves appear at $t\simeq 168$ ms and remain for $18.0$ ms.
The FT is characterized by a single peak that corresponds to the spatial period $\lambda\simeq 10$ $\mu$m, similar to the first calculation of maintaining the modulation depicted in Fig. \ref{fig: g1_7_fw}.
It is essential to note that Faraday waves appear a short time after turning off the modulation at $t_m=5$ ms, and the onset time of $168$ ms is delayed compared to that of the first calculation of the modulation.
This implies that the BEC retains the memory of the modulation even after it is turned off, and Faraday waves appear at $t\simeq 168$ ms.

\subsection{Appearance of Faraday waves with dissipation ($\gamma\neq0$)}
The dynamics with $\gamma$ yields significantly different results compared to that without $\gamma$.
When the BEC is subject to continuous modulation and dissipation with $\gamma=0.03$, the Faraday modes are excited at $t\simeq 360$ ms.
These waves remain for $24.8$ ms.
The density image and FT spectrum are \SI{similar to the result of the first calculation without $\gamma$ as shown in Fig. \ref{fig: g1_7_fw}.}
The FT is characterized by only one peak, which corresponds to the spatial period $\lambda\simeq 8.08$ $\mu$m.
%
%
The significant difference is found in the dynamics after the appearance of Faraday waves, which will be discussed in detail in the following section.
%
%

\section{Excitation of destabilization regime} \label{sec: Eme of non}
In this section, we consider the destabilization regime, which consists of several collective modes.
As described in the preceding section, there are two patterns for the appearance of the destabilization regime.
First, Faraday waves are excited initially, followed by other modes; eventually, the destabilization regime appears.
The other is caused when the injected energy is too strong; this regime appears suddenly, exciting several modes including the Faraday modes.
In this section, we focus on the first case in which Faraday waves are followed by the appearance of the destabilization regime. 
Interestingly, the dynamics for $\gamma\neq 0$ are completely different from that for $\gamma=0$.
For $\gamma\neq 0$, Faraday waves continue to appear and disappear.
%
%

\subsection{Dynamics of destabilization regime without dissipation ($\gamma=0$)} \label{sec: nonlinear_dis}
\begin{figure}[t] 
   \begin{center} 
       \includegraphics [width=1\columnwidth] {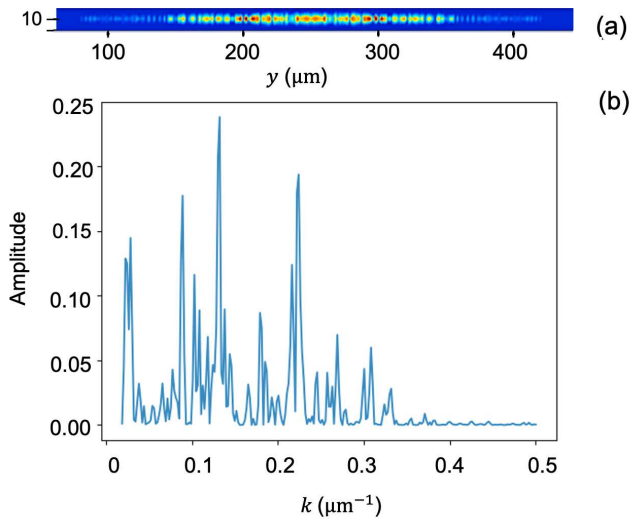}
       \caption{Appearance of destabilization regime for modulation at $t\simeq 100$ ms without $\gamma$. Driving frequency $\omega=1024\times 2\pi$ Hz and driving amplitude $\Delta B=5$ G. (a) Density image and (b) FT of one-dimensional density of destabilization regime.}
       \label{fig: g1_7_ins}
    \end{center}
\end{figure}

In the calculation for the modulation without $\gamma$, the destabilization regime appears at $t\simeq 88$ ms after the appearance of the Faraday waves.
Figure \ref{fig: g1_7_ins} (a) presents the density image of the destabilization regime, where the BEC cloud expands more along the $y$-axis than that depicted in Fig. \ref{fig: g1_7_fw} (a) at $t\simeq 80$ ms.
Figure \ref{fig: g1_7_ins} (b) depicts the appearances of multiple small peaks.
These peaks correspond to the excitation of other collective modes as well as the Faraday modes.
The animation presented in \cite{Sup1, Sup2} illustrates the time development of the density and FT from the Faraday waves depicted in Fig. \ref{fig: g1_7_fw} to the destabilization regime depicted in Fig. \ref{fig: g1_7_ins}.
%

%
\begin{figure}[t] 
   \begin{center} 
       \includegraphics [width=1.1\columnwidth] {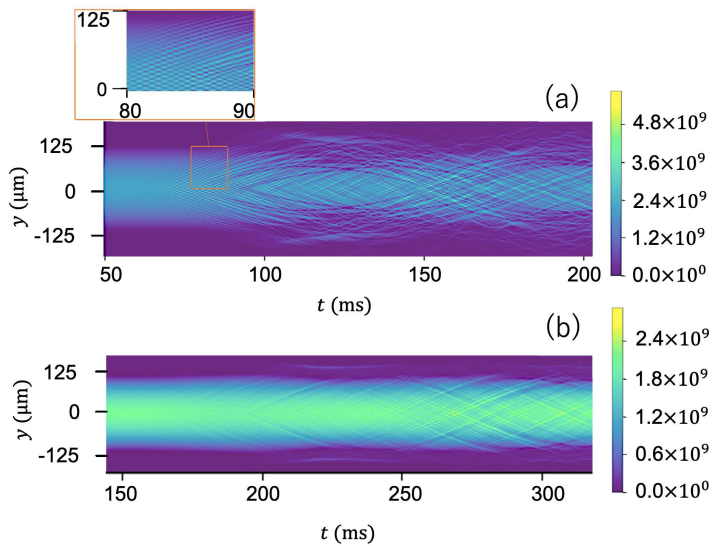}
       \caption{Time evolution of density at $\Delta B=5$ G and $\omega=2\omega_x$. Time evolution of density: (a) when modulation is continuously applied and (b) when modulation is turned off at $t_{m}=5$ ms without $\gamma$. Inset depicts appearance of Faraday waves.}
       \label{fig: fw_d_t}
    \end{center}
\end{figure}
\begin{figure*}[t!] 
    \includegraphics [width=2\columnwidth]{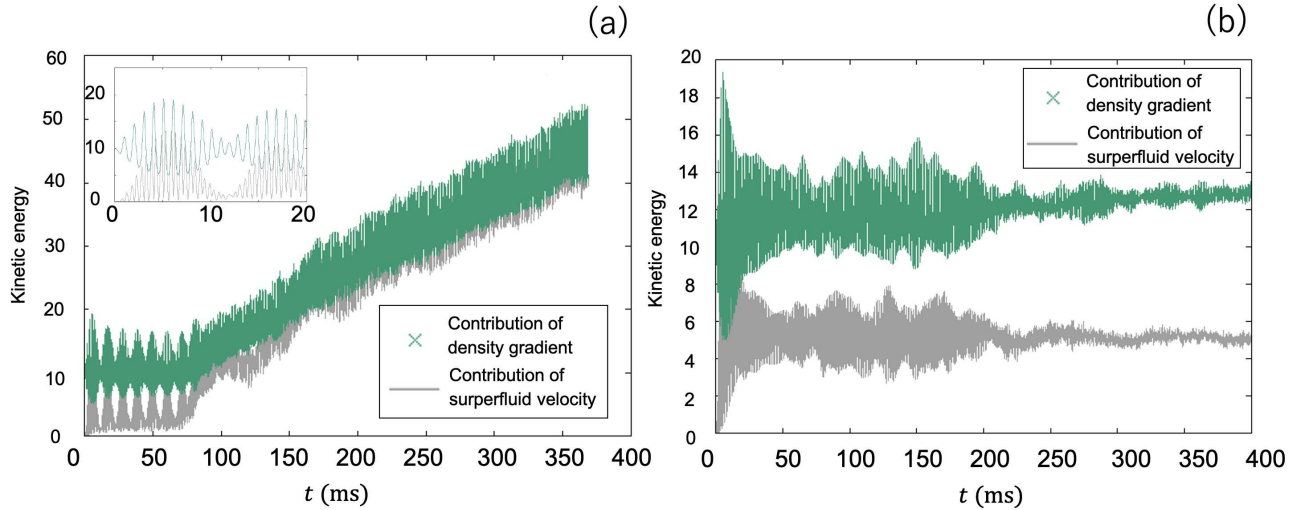}
    \caption{Kinetic energy dividing contributions to density gradient and superfluid velocity as depicted in Eq. (\ref{eq: Ek}). Driving frequency $\omega=1024\times 2\pi$ Hz and amplitude $\Delta B=5$ G. Green and gray plots represent the energy contributing to density gradient and superfluid velocity, respectively. Kinetic energy is calculated (a) for maintaining the modulation and (b) by turning the modulation off at $t_m=5$ ms without $\gamma$.}
    \label{fig: g1_Ek}
\end{figure*}
Figure \ref{fig: fw_d_t} presents the time evolutions of the density $|\psi|^2$.
For maintaining the modulation (Fig. \ref{fig: fw_d_t} (a)), several dips of the density move beyond the Thomas--Fermi radius ($\sim115$ $\mu$m) and are reflected by the trapping potential.
After that, the reflected dips return to the center of the BEC, move towards the opposite side, and are reflected again.
These movements have a periodicity of about $240$ ms.
These dips intersect with each other while maintaining their shapes.
This behavior is similar to that of dark solitons.
In the following section, we calculate the time evolution of a dark soliton and confirm that these dips are not dark solitons.
A similar behavior was reported in a previous study \cite{Balaz2014}, which addressed weakly inhomogeneous collisions and modulated the trapping potential.
Figure 5 of \cite{Balaz2014} depicts the expansion and shrinking of density dips.
Thus, the destabilization regime is not unique to the interaction modulation of our case.
Next, we depict the time evolution of the density in Fig. \ref{fig: fw_d_t} (b) when the modulation is turned off at $t_{m}=5$ ms.
The destabilization regime appears at $t\simeq 186$ ms; however, the expansion of the BEC is not clearly observed.
This regime does not necessarily accompany the expansion and suppression of the system, as seen when the modulation is maintained.
%

To investigate the difference between the first and second calculations, we calculate the kinetic energy by dividing it into contributions (Eq. (\ref{eq: Ek})) of the density gradient and the superfluid velocity, as depicted in Fig. \ref{fig: g1_Ek}.
The fast oscillations of the kinetic energy with a period of about $1$ ms correspond to the modulation of the interaction.
Figure \ref{fig: g1_Ek} (a) represents the kinetic energy while the modulation is maintained, and this energy causes quasiperiodic oscillations with periods of about $10$ and $1$ ms.
Faraday waves appear at $t\simeq 68$ ms, and the energy starts to increase at $t\simeq 75$ ms.
Thereafter, at $t\simeq 88$ ms, the BEC enters the destabilization regime.
The total energy continues to increase because of the modulation.
On the other hand, Fig. \ref{fig: g1_Ek} (b) indicates that the kinetic energy becomes statistically steady after turning off the modulation.
The kinetic energy causes quasiperiodic oscillations in this case too.
Faraday waves appear at $t=168$ ms, and the amplitude of the oscillations of the kinetic energy is reduced after $t\simeq 210$ ms.
The behaviors in the the first and second cases are significantly different; however, what happens is unclear, including the memory effect that causes the Faraday waves to appear a short time after the modulation is turned off.
%
%

%
\begin{figure*}[t!] 
       \includegraphics [width=2\columnwidth] {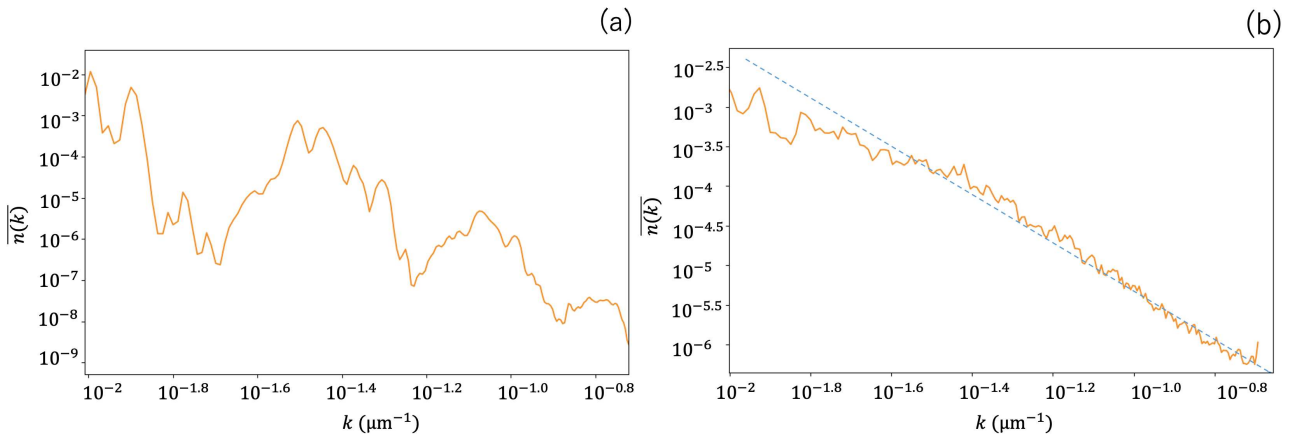}
       \caption{Time-averaged spectrum $\overline{n(k)}$ when (a) Faraday waves and (b) the destabilization regime appear. Driving frequency $\omega=2\omega_x$ and amplitude $\Delta B=5$ G. This calculation is performed by maintaining the modulation without $\gamma$. Orange solid line represents time-averaged $\overline{n(k)}$ for (a) $t\simeq 78.0\sim 87.0$ ms (b) $t\simeq 96.0\sim 165.2$ ms. Blue dashed line represents fitting function $\overline{n(k)}\propto k^{-3.09}$.}
       \label{fig: int_psi_k}
\end{figure*}
Figure \ref{fig: fw_d_t} (a) illustrates that several dips of density move around in the BECs and induce the destabilization regime.
To study whether the destabilization regime is characterized by dark solitons, we compare the dip of the density with that of a dark soliton \cite{Pethick2008}.
When the dark solitons have a density $n_0$ in a stationary state and the minimum density value $n_{min}$, their velocity $u$ is expressed as follows:
\begin{equation} \label{eq: xi_u}
u=s \sqrt{\frac{n_{min}}{n_0}},
\end{equation}
with the sound velocity $s=\sqrt{n_0g/m}$.
The following phase difference exists across the dark solitons:
\begin{equation} \label{eq: density of soliton}
\Delta\phi=-2\cos^{-1}\left(\frac{u}{s}\right).
\end{equation}
The ratio $n_{min}/n_0$ is approximately $0.001$ and $s=340$ $\mu$m$/$ms for the dips in our simulation.
If the dips are dark solitons, the soliton velocity should be approximately $11$ $\mu$m$/$ms; however, this velocity is larger than approximately $2$ $\mu$m$/$ms for the dips depicted in Fig. \ref{fig: fw_d_t} (a).
Dark solitons are expected to reflect a phase difference of approximately $-\pi$, which is not observed for the dips in our simulation.
Therefore, we conclude that these dips do not represent dark solitons.
%

Next, we calculate the spectrum $n(k)=|\psi_k|^2$ \SI{to characterize the destabilization regime.}
\SI{Figure \ref{fig: int_psi_k} (a) shows the time-averaged $\overline{n(k)}$ for the early period of $t\simeq 78.0\sim 87.0$ ms while several modes including the Faraday waves are excited.}
\SI{Figure \ref{fig: int_psi_k} (b) shows the time-averaged $\overline{n(k)}$ for the late period of $t\simeq 96.0\sim 165.2$ ms while the dynamics enters the destabilization regime.}
The spectrum illustrates the power law $\overline{n(k)}\propto k^{-3.09}$.
The exponent $-3.09$ is similar to the power \SI{$-3.0$} of the typical wave turbulence \SI{\cite{Zakharov1992}}.
The spectrum $\overline{n(k)}$ deviates from the power law at low wave numbers owing to the finite size effect.
The system size corresponds to $\log{k}=-2.2$.
The detailed studies of the turbulence and the exponent are beyond the scope of this paper.

\subsection{Dynamics of destabilization regime with dissipation ($\gamma\neq 0$)} \label{sec: nonlinear_dis}
\begin{figure}[t] 
   \begin{center} 
       \includegraphics [width=1.1\columnwidth] {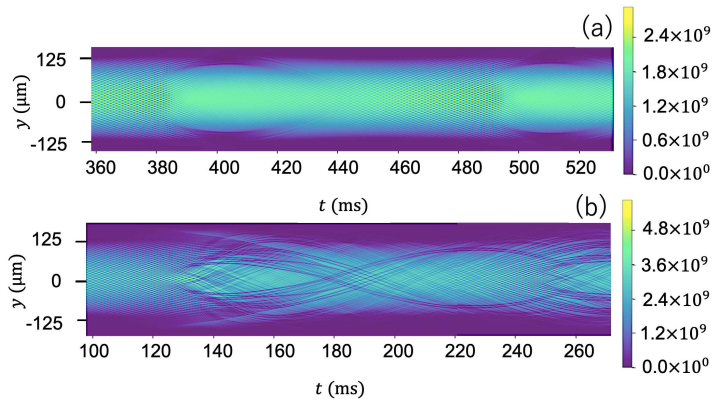}
       \caption{Time evolution of density at $\Delta B=5$ G and $\omega=1024\times 2\pi$ when modulations with (a) $\gamma=0.03$ and (b) $\gamma=0.01$ are maintained}
       \label{fig: dis_d_t}
    \end{center}
\end{figure}

The calculations without $\gamma$ indicate that the BEC enters the destabilization regime after the appearance of Faraday waves.
However, the calculation with $\gamma=0.03$ indicates that Faraday waves repeat the suppression and growth.
The time development of these dynamics is presented in Fig. \ref{fig: dis_d_t} (a).
As illustrated, Faraday modes are excited at $t\simeq 360$ ms, and they disappear at $t\simeq 385$ ms.
The BEC is stationary during the disappearance of these waves.
Thereafter, the waves reappear at $t\simeq 456$ ms and disappear at $t\simeq 492$ ms.
Subsequently, the suppression and growth of these waves are repeated several times with approximately the same period.
Similar revival and suppression phenomena have also been reported experimentally \cite{Nguyen2019}.
Our simulations demonstrate that dissipation is indispensable for such phenomena.
On the other hand, the result obtained for $\gamma=0.01$ indicates a significantly different behavior.
Faraday waves appear at $t\simeq 85$ ms, and the destabilization regime appears at $t\simeq 125$ ms.
Interestingly, Faraday waves and the destabilization regime repeat the revival and suppression, as depicted in Fig. \ref{fig: dis_d_t} (b).
That is, Faraday waves and the destabilization regime reappear at $t\simeq 230$ ms and $t\simeq 250$ ms, respectively.
It should be noted that the reappearance of Faraday waves is accompanied by weak excitations of collective modes owing to the depression of the destabilization regime.
%

%
\begin{figure*}[t!] 
    \includegraphics [width=2\columnwidth]{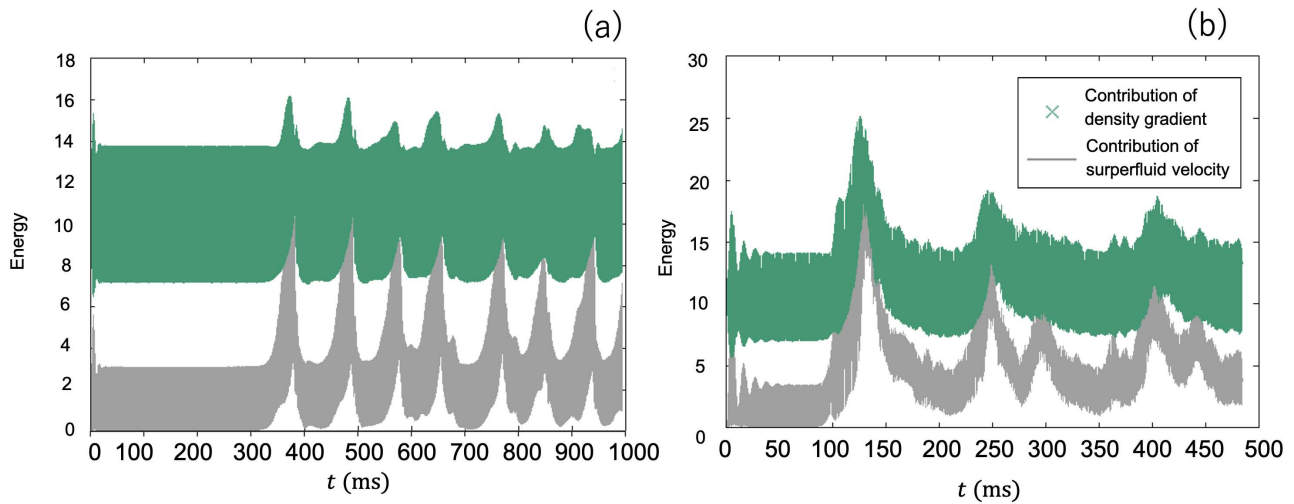}
    \caption{Kinetic energy dividing contributions to density gradient and superfluid velocity as depicted in Eq. (\ref{eq: Ek}). Driving frequency $\omega=1024\times 2\pi$ Hz and amplitude $\Delta B=5$ G. Green and gray plots represent energy contributing to density gradient and superfluid velocity, respectively. Kinetic energy is calculated by maintaining modulation with (a) $\gamma=0.03$ and (b) $\gamma=0.01$.}
    \label{fig: dis_Ek}
\end{figure*}
To clarify the difference in the dynamics with and without $\gamma$, we show the time development of the kinetic energy.
Figure \ref{fig: dis_Ek} (a), for $\gamma=0.03$, indicates that the kinetic energy owing to the density gradient starts to increase at $t\simeq 320$ ms and decrease at $t\simeq 360$ ms; thereafter, Faraday waves appear.
At $t\simeq 460$ ms, this energy increases again.
In this case, the dynamics does not lead to quasiperiodic oscillations as depicted in Fig. \ref{fig: g1_Ek} (a), and the kinetic energy is smaller than that in the case without $\gamma$.
A comparison between Figs. \ref{fig: dis_d_t} (a) and \ref{fig: dis_Ek} (a) indicates that the oscillations of the kinetic energy correspond to the appearance and disappearance of Faraday waves, which is caused by the balance between the energy of injection and dissipation.
Figure \ref{fig: dis_Ek} (b), for $\gamma=0.01$, indicates that the kinetic energy increases because $\gamma$ is smaller than that depicted in Fig. \ref{fig: dis_Ek} (a).
The dynamics for $\gamma=0.03$ and $\gamma=0.01$ are completely different; however, their injection and dissipation energies are balanced in both cases.
%

When $\gamma=0.03$ and $\Delta B=15$ G, significantly more energy is injected into the BEC, and the destabilization regime appears at $t\simeq 39.8$ ms without Faraday waves.
In this case, we cannot observe the suppression and growth of Faraday waves (similar to the nondissipative cases) because the injected energy is very large, implying that the balance between the injection and dissipation energies cannot be maintained.
%

The calculation with $\gamma=0.03$ is different from the nondissipative cases in three ways.
First, the dissipation suppresses the excitations of the collective modes.
At this time, the Faraday modes are also suppressed; however, they appear because they are most strongly excited against the dissipation, as depicted in Fig. \ref{fig: dis_d_t} (a).
That is, the dissipation prevents the appearance of the destabilization regime but not the revival of the Faraday waves.
Second, the onset time of the Faraday waves with $\gamma$ is delayed compared to that without $\gamma$.
This is because the injected energy is swept away by the dissipation, and it takes more time to excite Faraday waves.
Finally, the peak of the FT for $\gamma=0.03$ is higher than that for $\gamma=0$, implying  that Faraday waves are more clearly observed for $\gamma=0.03$ than for $\gamma=0$.
The reason for this phenomenon is unknown.

\subsection{Dynamics of Faraday waves and destabilization regime} \label{sec: dynamics}
In this section, we summarize the dynamics of Faraday waves and the destabilization regime.
The summary \SI{of the numerical conditions} is presented in Table \ref{table:data_type}.
When we maintain the modulation without $\gamma$, the density of the BEC expands and shrinks continuously while maintaining the pancake-shape as that before the appearance of the Faraday waves.
During this time, the kinetic energy produces a quasiperiodic motion, as depicted in Fig. \ref{fig: g1_Ek} (a).
After tens of milliseconds, the kinetic energy owing to the density gradient increases and the Faraday modes are excited, as depicted in Fig. \ref{fig: g1_7_fw}.
During the excitation of these modes, the suppression and revival of the Faraday waves are repeated with a period of a few milliseconds, as presented in Fig. \ref{fig: g1_7_fw_dy}.
Next, these waves are suppressed completely and the condensate expands beyond the Thomas--Fermi radius, as depicted in Fig. \ref{fig: fw_d_t} (a).
Thereafter, the potential energy increases, as does the total energy.
Therefore, other collective modes are excited and the dynamics enters the destabilization regime, which accompanies the expansion and suppression of the BEC.
If we continue the excitation, the BEC may eventually be suppressed and the system goes beyond what the GP mean-field can describe.
On the other hand, the calculation upon turning off the modulation at $t_m=5$ ms without $\gamma$ indicates the appearance of Faraday waves and the destabilization regime similar to Fig. \ref{fig: fw_d_t} (b).
There are three differences associated: the first is the onset time of the instability.
The onset time when the excitation is turned off is delayed as compared with that when the modulation is maintained.
The second difference is that the kinetic energy caused by the density gradient converges.
The third is that the destabilization regime does not clearly indicate the expansion and suppression of the BEC.
%

%
\begin{table*}[t!]
    \caption{Modulation method and results}
    \label{table:data_type}
    \hspace*{-1cm}
    \begin{tabular*} {2.25\columnwidth}{|c|c|c|c|} \hline
        Modulation method & After applying modulation & After appearance of Faraday waves \\ \hline
        Maintaining modulation ($\gamma=0$) & Appearance of Faraday waves & Appearance of destabilization regime \\ \hline
        Turning off modulation ($\gamma=0$) & Same as above & Same as above \\ \hline
        Maintaining modulation ($\gamma=0.03$) & Same as above & Revival and suppression of Faraday waves \\ \hline
        Maintaining modulation ($\gamma=0.01$)  & Same as above & Revival and suppression of Faraday waves and destabilization regime \\ \hline
    \end{tabular*}
\end{table*}

The calculation while maintaining the modulation with $\gamma=0.03$ reveals the revival and suppression of the Faraday waves without the destabilization regime.
The energy between the injection and dissipation is balanced, and the kinetic energy oscillates corresponding to the appearance of the Faraday waves.
In addition, the revival and suppression of the Faraday waves and the destabilization regime repeat for the calculation with $\gamma=0.01$.
Thus, it is clear that the calculations without $\gamma$ have different dynamics as compared to those with $\gamma$.
Because the realistic system encounters certain dissipations, it should behave similar to the system with dissipation described in this paper.

\section{Conclusion} \label{sec: Con}
We calculated a GP equation to investigate the dynamics of Faraday waves and the destabilization regime in a pancake-shaped BEC by modulating the interaction.
\SI{We found the characteristic nonlinear dynamics of the BEC with or without the dissipation.}
\SI{For the nondissipative cases, after the first appearance of the Faraday waves, the destabilization regime, which consists of collective modes, appears.}
When the destabilization regime appears, the BEC expands and shrinks, which is accompanied by dips in the densities intersecting each other, while maintaining their shapes, similar to dark solitons.
However, these dips do not represent dark solitons because their profiles are significantly different.
Furthermore, for $\Delta B=5$ G and $\omega=2\omega_x$, the BEC develops into wave turbulence.
%
%
%
%

\SI{The simulation show that the dissipation is necessary for the revival and the suppression of Faraday waves reported in the previous experiments \cite{Nguyen2019}.}
The simulation with the weak dissipation reflects the appearance of the destabilization regime after the excitation of the Faraday modes.
\SI{These excitations appear and disappear repeatedly.}
\SI{This simulation reveals the nonequilibrium state sustained by the balance between the injection and the dissipation.}
%

In this study, we investigated the case of modulating the interaction in the pancake-shaped BEC.
The appearance and the dynamics of the Faraday waves depend on the geometry of the BEC, and it is a future work to study the Faraday waves in other geometries of the BEC, \SI{for example, what patterns can be observed by changing the geometry from one-dimensional system to two-dimensional.}
However, another means of exciting Faraday waves is modulating the trapping potential \cite{Engels2007}.
If a BEC is modulated by the trapping potential, we expect similar dynamics.
Furthermore, another interesting theme is the dynamics of Faraday waves in binary BECs.
In this case, the dynamics are investigated by modulating the interaction or the potential.
When $\gamma=0$, binary BECs are expected to behave similar to that observed in previous studies \citep{Balaz2012, Maity2020}, and the dynamics may generate the destabilization regime after the appearance of the Faraday waves.
The system with $\gamma\neq0$ is realistic but is not studied numerically in the previous works.
This study of Faraday waves reflects a good platform for pattern formulation in a BEC.

M.T. acknowledges support from JSPS KAKENHI (Grant Number JP20H01855).
We would like to thank Sousuke Inui for helping the analysis used in this work.

\bibliography{research.bib}

\end{document}